# Teaching cloud computing: a software engineering perspective

Ian Sommerville, School of Computer Science, St Andrews University, Scotland.

When a new area of a discipline emerges, university teachers are faced with the problem of when (and if) this should be introduced into their courses. They have to ask two questions: firstly, is this something that is attracting a lot of publicity but which will have little long-term impact and secondly, what are the essential elements of this topic, which should be part of an enduring course. Cloud computing is no exception here although I think that it is fair to say that the first question has already been answered. Cloud computing is not going to go away and we certainly do have a responsibility to include this in our courses.

We have already seen a number of cloud computing courses being offered and even some Master's degrees. Lee Gillam's course at Surrey University was one of the earliest and he has written some reflections on this course[1]. His course is a general introductory course in cloud computing that starts by introducing the notions of software, platform and infrastructure as a service. It covers relevant cloud protocols such as SOAP and REST and discusses the map-reduce computational model and its instantiation in Hadoop. The course makes comparisons with grid and peer to peer computing and discusses service-level agreements, cloud economics and security.

In this article, I want to avoid simply listing what I think should be included in a cloud computing course. Rather, I want to look at the topic of teaching cloud computing from a more abstract perspective and discuss the issues and problems of teaching this subject. In line with the theme of this issue, I will look at this from a software engineering rather than a business or information systems perspective – teaching cloud computing in these areas would be rather different from what I discuss here.

When considering issues of teaching, I find it helpful to consider topics from three perspectives:

1. *Sensitisation*  Telling students about something and how it is used. Essentially, the aim here is to ensure that they are not surprised when they encounter this when they leave university. There is no expectation that students will have practical skills or theoretical knowledge. Typically, sensitization is the first stage of introducing a new topic into the curriculum.

2. *Practice*  At this stage, students are given some tuition in the practical elements of a topic – so, in cloud computing, they may be asked to provision some servers on a cloud service such as AWS or Microsoft Azure. Practice is usually the next stage after sensitization.

3. *Principles*  At this stage, we are trying to abstract the fundamental principles of a topic and present these to students. An understanding of these principles means that students

---

[1] Lee Gillam, Bin Li and John O'Loughlin. "Teaching Clouds: Lessons Taught and Lessons Learnt ". In *Cloud Computing for Teaching and Learning: Strategies for Design and Implementation*. Ed. Lee Chao. IGI Global, 2012.

> have general rather than specific knowledge which can then be applied to understanding new systems and services.

The time at which coverage of a topic is included in a course very much depends on the underlying course philosophy. Some courses are very practically oriented so will introduce cloud computing practice at an early stage; other courses are more theoretical and will delay the introduction of practical work until there is a firm body of principles on which this can be based.

For cloud computing, for sure we have reached the stage of sensitization and I believe that something on this topic should be included in all courses. All computer science and software engineering students are already cloud users, even if they don't understand this explicitly, so it makes sense to introduce the cloud early in the course. In my university, this is in first year where students have a short three hour introduction to cloud computing plus coursework that requires them to do further reading in the area.

The introduction covers the development and evolution of cloud computing, including a discussion of how services such as iTunes are based on a cloud framework. We then briefly discuss some of the technical background – virtualization and the notions of infrastructure, platform and software as a service. Coursework requires them to do further reading around issues such as reliability, security and privacy. We do not, at this stage, introduce cloud-based practical work.

This is the easy bit of teaching cloud computing. The next stage is to make a decision on whether there are significant educational benefits of introducing a practical course in cloud computing. In my view, there is very little educational merit in teaching students how to provision servers in an IaaS service, such as Amazon Web Services. There is a lot of detailed specific knowledge here but this is not readily transferable to other providers and, for sure, it will become redundant very quickly as companies such as Rightscale introduce new cloud management tools.

However, the value of using IaaS in a software engineering course is that it makes it technically very simple to support server-based project work, which without the cloud, may have required access to dedicated hardware. Servers can be set up for individual students and student groups for the duration of the project without incurring capital costs for hardware which has a relatively low level of utilization. As part of a project introduction, it may well be appropriate to spend a few hours introducing this topic and helping students navigate through the large volume of tutorial material that is available.

The problems here are likely to be managerial rather than technical. The charging system used by cloud providers is usually based on a credit card and staff involved in teaching may not be authorized to use university credit cards. If students are allowed to provision servers themselves, they may unwittingly set up expensive server types or leave servers running when they are not required. These result in additional costs and the cost risk may not be seen as acceptable by the institution.

Where practical work becomes much more interesting (I think) is when we consider PaaS and student programming assignments are based on a cloud platform such as Google's App Engine. This allows us to introduce some of the fundamental differences between general programming and programming the cloud such as managing scaleability, multi-tenancy and schema-free databases.

The problem that many universities face here is the amount of time they have available in their courses. PaaS interfaces are quite complex and involve new concepts for students so a significant time has to be spent introducing these and dealing with the details of a specific interface. The concepts are certainly transferable across providers but there is always the danger that students will become immersed in the detail of one PaaS platform without understanding the generalities of what they are doing.

It is when we reach the final stage – that of principles, where I think syllabus design becomes particularly difficult. What are the fundamental enduring principles of cloud computing that are distinct from more general principles of distributed computing? Is the cloud simply a special case of a distributed computing system? Many academics will argue that this is the case but I believe the key distinction that we get from considering cloud computing in its own right is that it allows us to address issues of scale.

Teaching about scale has been a perennial problem in software engineering. As a discipline, this has evolved to address these problems and discusses the methods and tools to support the creation of large, complex systems. Yet, by and large, students never get any experience of large systems and so often fail to appreciate the value of software engineering methods. Whilst the cloud does not help with illustrating the socio-technical problems that arise with large systems, it does provide a vehicle for experimentation with the technical issues of very large-scale systems.

This is now possible because governments are starting to release large-scale data sets of public information. For example, in the UK, the government has released more than 8000 data sets[2] including data on health, science environment, crime and education. This presents us with an unprecedented opportunity to educate students in how to use the cloud for large-scale data processing.

A course on cloud software engineering should be a senior or a graduate course - the key topics and principles that might be covered are:

1. *The map-reduce paradigm for independent computation*. Using Hadoop to write map-reduce programs. Issues and problems with this approach and, in particular, the problems of using it with transactional systems and relational databases.

2. *Schema-free databases and their use*. A discussion of the problems with map-reduce leads naturally to this topic. Areas where schema-free databases are appropriate and applications for which they are inappropriate.

3. *Service-oriented computing*. Arguably, this is really part of a more general distributed computing course but services and cloud computing have become synonymous. Topics covered here should include RESTful and 'big web services', decomposing a system into services,

4. *Multi-tenancy*. This leads naturally from a discussion of service-oriented systems and how services are evolving and should focus on service design so that single instances of the service are shared with data from multiple customers being held in the same database.

5. *Security and compliance*. This is a topic that has to be included although it is quite difficult to approach it from a principled perspective. Many of the issues of cloud security are simply general security issues which are equally applicable to self-hosted systems. Compliance requirements are very significant indeed but vary significantly from one country to another.

6. *Design for resilience*. Many people make the assumption that all you have to do is to set up a system in the cloud and everything else is done for you by the cloud provider. Nothing could be further from the truth and it is important to discuss how to design cloud-based systems with redundancy within and across cloud providers so that they can tolerate and quickly recover from provider failures.

Ideally, students would come to this course having gained some practical experience of cloud programming using some PaaS system. Realistically, however, the pressure on practical work in computer science and software engineering courses is such that this is probably an unrealistic pre-requisite and the practical work associated with the course would have to cover the fundamentals of PaaS before moving on to programming with Hadoop and service implementation. I don't think it matters much which PaaS platform is used and the choice may be governed by the other material in the course (e.g. if students have experience of .Net, then Microsoft Azure is the logical choice; if they have experience of Python, Google App Engine is most appropriate, etc.).

---

[2] http://data.gov.uk

I do not advocate the teaching of topics such as service-level agreements, cloud economics, public versus private clouds, etc. These may be appropriate in a cloud computing for business course but there are few principles embodied in these topics and the associated knowledge will become out of date very quickly.

The most significant barrier to the introduction of high-quality cloud computing courses at the moment is the current skill base of university staff. The majority of faculty members were appointed before cloud computing became available and have little personal research or teaching experience in this area. Faculty are increasingly pressurized to focus on research achievement and it is very difficult for them to find time to explore new areas for their teaching. Even when there are a few faculty with experience and interest in this area it is sometimes difficult to get new courses approved simply because the majority of staff do not understand how the work relates to computer science or software engineering as they see it.

In conclusion, at one level, the cloud is nothing new for software engineering. The fundamental issues of managing problem and solution complexity remain the same and it will always be very difficult for students to understand and appreciate these. However, the cloud means that we can at least partially address the problems of teaching about scale and we do have a responsibility to ensure that our students are aware of and can make use of modern techniques of system implementation. What is important here is that we think carefully about the material to be included in courses so that it is principled, with lasting value and is not simply an ephemeral reflection of the features that are currently available from some cloud provider.

Finally, let me say a word on the introduction of degrees and Masters courses on cloud computing. I think that these are a cynical re-badging of existing courses in the hope of attracting students to a high-profile topic. There is simply not enough fundamental material at the moment (and I suspect that there may never be) to justify a degree course in this topic in its own right.

**Teaching materials**

All of the major cloud providers are interested in having their systems covered in university courses and provide free teaching materials. Examples of material that is available include:

Windows Azure Training Kit.
http://www.microsoft.com/en-gb/download/details.aspx?id=8396

Amazon Web Services tutorials.
http://aws.amazon.com/articles/

Google App Engine.
http://developers.google.com/appengine/docs/

IBM Cloud computing.
http://www.ibm.com/developerworks/university/teachingtopics/cloud_computing.html